# "Water-cycle" mechanism for writing and erasing nanostructures at the LaAlO$_3$/SrTiO$_3$ interface

Feng Bi[1], Daniela F. Bogorin[1], Cheng Cen[1], Chung Wung Bark[2], Jae-Wan Park[2], Chang-Beom Eom[2], Jeremy Levy[1*]

[1]Department of Physics and Astronomy, University of Pittsburgh, Pittsburgh, PA 15260, USA

[2]Department of Materials Science, University of Wisconsin, Madison, WI 53706, USA

*jlevy@pitt.edu

**Abstract:**

Nanoscale control of the metal-insulator transition in LaAlO$_3$/ SrTiO$_3$ heterostructures can be achieved using local voltages applied by a conductive atomic-force microscope probe. One proposed mechanism for the writing and erasing process involves an adsorbed H$_2$O layer at the top LaAlO$_3$ surface. In this picture, water molecules dissociates into OH$^-$ and H$^+$ which are then selectively removed by a biased AFM probe. To test this mechanism, writing and erasing experiments are performed in a vacuum AFM using various gas mixtures. Writing ability is suppressed in those environments where H$_2$O is not present. The stability of written nanostructures is found to be strongly associated with the ambient environment. The self-erasure process in air can be strongly suppressed by creating a modest vacuum or replacing the humid air with dry inert gas. These experiments provide strong constraints for theories of both the writing process as well as the origin of interfacial conductance.

The discovery of a high mobility quasi-two-dimensional electron gas (q-2DEG) at the LaAlO$_3$/ SrTiO$_3$ (LAO/STO) heterointerface[1] has drawn great interest towards its transport property [2-5], potential devices application [3,6-8] and its physical mechanism [9-14] . One defining characteristic of this family of heterostructures is the abrupt transition from an insulating to conducting interface for $n \geq 4$ unit cells [5]. (We define $n$ uc-LAO/STO to refer to $n$ unit cells of LaAlO$_3$ grown on TiO$_2$-terminated SrTiO$_3$.) To explain this transition, a number of mechanisms have been offered including electronic reconstruction (sometimes referred to as "polar catastrophe") [9], structural deformations [14], unintentional or intrinsic dopants [10], interfacial intermixing [12] and oxygen vacancies [11]. Theoretical investigations have predicted a critical thickness ranging from $n$=3-4 uc [15]. For 3uc-LAO/STO structures, reversible nanoscale control of the metal-insulator transition was reported [3]. Positive voltages applied to a conductive atomic force microscope (c-AFM) probe in contact with the top LAO surface produce local conducting regions at the LAO/STO interface; negative voltages restore the interface to its initial insulating state. The process was found to be repeatable over hundreds of cycles [6], effectively ruling out an early theory involving the formation of oxygen vacancies at the top LAO surface [3]. Conducting regions were found to be stable under atmospheric conditions for ~ 1 day, and indefinitely under vacuum conditions [6].

A physical understanding of the writing and erasing mechanism is important for fundamental reasons and also for the development of future technologies that are based on the stability of these nanostructures. Conducting islands with densities >150 Tb/in$^2$ have been demonstrated [3], and transistors with channel lengths of 2 nm have been reported [6]. Such an understanding can help in the development of conditions that can stabilize these structures over time scales that are relevant for information storage and processing applications (i.e., ~10 years).

One possible mechanism for the writing process involves adsorbed $H_2O$ which dissociates into $OH^-$ and $H^+$ on the LAO surface. First principles calculations [16] show $H_2O$ binds strongly to the $AlO_2$ outer surface at and below room temperature, and dissociates into $OH^-$ and $H^+$ adsorbates. During the writing process, the positively biased AFM probe removes some of the OH adsorbates, thus locally charging the top surface with an excess of $H^+$ ions. This charge writing, which has also been observed for bulk $LaAlO_3$ crystals [17], acts to modulation dope the LAO/STO interface, switching it from insulating to conducting. During the erasing process, the negatively charged AFM probe removes H adsorbates, restoring the $OH^-$ - $H^+$ balance, and the interface reverts back to an insulating state. We refer to this process as a "water cycle" because it permits multiple writing and erasing without physical modification of the oxide heterostructure.

Here we investigate the writing and erasing process on 3uc-LAO/STO heterostructures under a variety of atmospheric conditions, in order to constrain physical models of the writing and erasing procedure and the origin of the interfacial electron gas. Thin films (3 u.c.) of $LaAlO_3$ were deposited on a $TiO_2$-terminated (001) $SrTiO_3$ substrates by pulsed laser deposition with *in situ* high pressure reflection high energy electron diffraction (RHEED) [18]. Growth was at a temperature of 550°C and $O_2$ pressure of $1\times10^{-3}$ Torr.

After growth, electrical contacts to the interface were prepared by milling 25nm deep trenches via an Ar-ion mill and filling them with Au/Ti bilayer (2nm adhesion Ti layer and 23nm Au layer).

**To perform c-AFM experiments, a vacuum AFM (FIG. 1(a)) is employed that is capable of operation down to $10^{-5}$ Torr and allows controlled introduction of various gases. Writing and erasing experiments (FIG. 1**(b,c)) are performed under a variety of conditions. We use the parameter $V_{tip}$=10V, litho speed= 500nm/s for writing and $V_{tip}$=-10V, litho speed=10nm/s for erasing. The conductance of nanostructures is monitored in real time by a lock-in amplifier as the ambient gaseous environment is modified in a controlled fashion. Here conductance is defined as $G=I_{RMS}/V_{RMS}$. A sinusoidal voltage (amplitude 0.4 $V_{RMS}$, frequency 23 Hz) is applied to one electrode, and the resulting current $I_{RMS}$ from the second electrode is measured by the lock-in amplifier. The pressure in the experiment is monitored using an ion gauge, which operates over a range Atmosphere-$10^{-9}$ Torr, with an accuracy ±15% <100 mbar and ±30% below $10^{-3}$ mbar. Prior to the writing the LAO surface is raster-scanned twice with $V_{tip}$ = -10 V and $V_{tip}$=+10V alternatively, to remove any adsorbates on the LAO surface.

A straightforward test of the water cycle mechanism outlined above replaces atmospheric conditions with gas environments that lack $H_2O$. **FIG.** 2 shows the results of a number of writing experiments performed using dry air (**FIG.** 2(a)), helium gas (**FIG.** 2(b)), and dry nitrogen (**FIG.** 2(c)) under pressures ranging from $10^{-2}$-$10^2$ Torr. Nanowires were not formed under any of these conditions. To verify that the sample was not adversely affected during these experiments, the sample was subsequently exposed to air (28% RH) and a nanowire was written with ~120 nS conductance (**F**IG. 2(d)). The nanowire was then erased and the AFM was evacuated to base pressure ($1.8 \times 10^{-5}$ Torr). Under vacuum conditions, it was again not possible to create conducting nanostructures.

We also check the ability to erase nanostructures under vacuum conditions. A nanostructure is created under atmospheric conditions, and the AFM is evacuated to base pressure. After that, the conductance of such nanostructure is stabilized around 20 nS. **F**IG. 2(d) inset illustrates that erasure is still achievable under vacuum conditions.

Finally, we illustrate how the process of self-erasure depends on atmospheric conditions (FIG. 3). Self-erasure process of a single nanowire is observed in air and vacuum subsequently (FIG. 3(a)). A nanowire written at atmosphere pressure (RH=28%) exhibits a rapid initial decay. At $t=300$ s after writing the nanowire, the system is evacuated, and then reaches a pressure of $1.7 \times 10^{-4}$ Torr. During this time, the nanowire conductance quickly stabilizes and reaches a constant value. At $t=1670$ s the system is vented, and the nanowire conductance resumes its decay. This experiment demonstrates that self-erasure is directly associated with atmospheric conditions, and that it can be slowed significantly or halted under modest vacuum conditions $\sim 10^{-3}$ Torr. These results are consistent with previous observations [6]. FIG. 3(b) compares the self-erasure processes of a nanowire kept in air with a nanowire kept in 1atm dry nitrogen gas. The red curve in FIG. 3(b) shows that a naowire is written in air and then decay quickly in such environment (31% relative humidity air). It takes about only 2.2 hours for this nanowire to decay close to the background. For the blue curve, the nanowire is first formed in air and then the system is evacuated to $1.4 \times 10^{-4}$ Torr. At t=1900s, 1 atm dry nitrogen gas is introduced to the system. Compared with in humid air, the self-erasure process in 1 atm dry nitrogen gas is strongly suppressed. The nanowire conductance decays by a factor of 50 or so over a 72 hour period, but remains a nanowire as evidenced by the subsequent erasure. This behavior is comparable to what was reported for vacuum conditions in Ref. [6]. Therefore, these results illustrate that vacuum conditions are not required for long-term retention of nanoscale structures, although some self-erasure is evident.

The experiments described above constrain not only models for the writing and erasing of nanostructures at LaAlO$_3$/SrTiO$_3$ interfaces; they also constrain models of the interfacial conductivity itself. It is difficult to formulate a physical model that invokes only intermixing or oxygen diffusion—either at the top LaAlO$_3$ surface or at the interface itself—to explain the interfacial conduction. Rather, the interfacial conductance results directly from modulation doping of an otherwise insulating interface, which is stabilized by the screening of the polar discontinuity between LaAlO$_3$ and SrTiO$_3$.

## Acknowledgements

Work at the University of Wisconsin was supported by funding from the National Science Foundation (DMR-0906443) and the DOE Office of Basic Energy Sciences (DE-FG02-06ER46327). Work at the University of Pittsburgh was supported by the National

Science Foundation (DMR-0704022), DARPA seedling (W911NF-09-1-0258) and the Fine Foundation. We gratefully acknowledge stimulating discussions with C. S. Hellberg.

# Figures

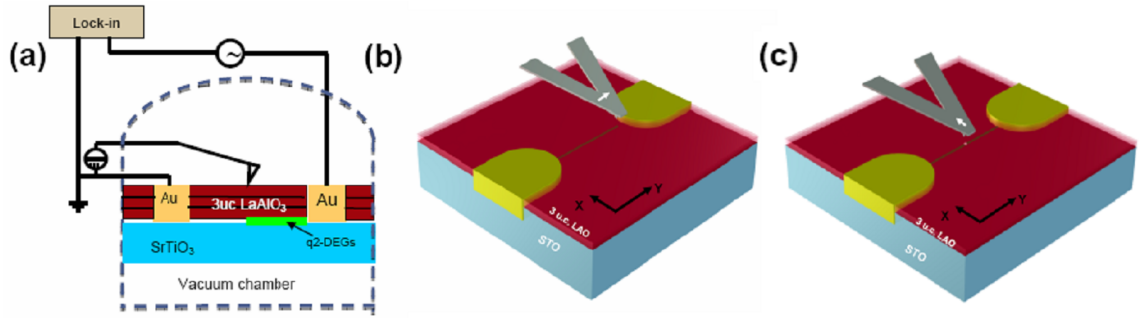

FIG. 1. Writing and erasing nanowires at 3u.c. LaAlO$_3$/SrTiO$_3$ interface. (a) Side view schematic illustration about how a conducting AFM probe writes a nanowire. (b) Top view schematic of a writing experiment in which a nanowire is created with a positive biased tip. (c) Top view schematic of a cutting experiment in which a nanowire is locally erased with a negatively biased tip.

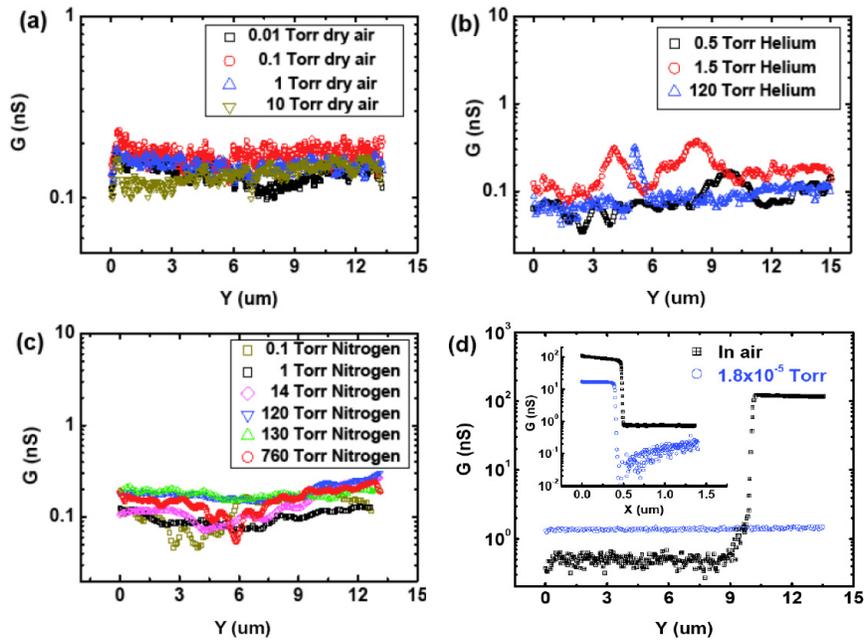

FIG. 2. Nanowire writing under various atmospheric conditions. Writing a nanowire in (a) air, (b) helium, and (c) nitrogen environments. (d) Subsequent writing nanowires under vacuum and atmospheric conditions confirms that no irreversible changes have occurred to the sample. Inset shows erasing under vacuum conditions, illustrating that the erasure process is insensitive to atmospheric conditions.

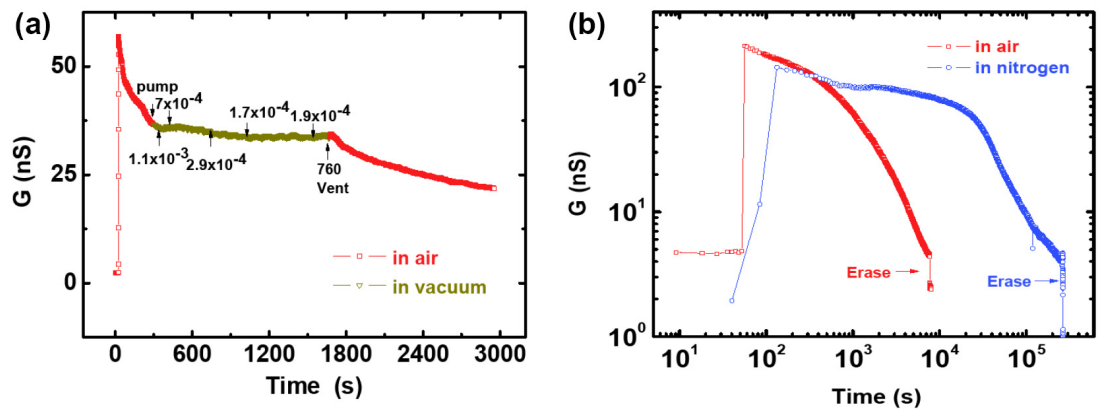

FIG. 3. Experiment showing effect of atmosphere on nanostructure self-erasure. (a) The self-erasure process of a single nanowire in air, vacuum and back to air subsequently. Unit of pressure in (a) is Torr. (b) Compare the self-erasure processes of one nanowire in air (RH=31%) and another nanowire in 1atm dry nitrogen gas.

# References

[1]   A. Ohtomo and H. Y. Hwang, Nature **427** (6973), 423 (2004).
[2]   A. Ohtomo and H. Y. Hwang, Nature **441** (7089), 120 (2006).
[3]   C. Cen, S. Thiel, G. Hammerl, C. W. Schneider, K. E. Andersen, C. S. Hellberg, J. Mannhart, and J. Levy, Nat. Mater. **7** (4), 298 (2008).
[4]   N. Reyren, S. Thiel, A. D. Caviglia, L. F. Kourkoutis, G. Hammerl, C. Richter, C. W. Schneider, T. Kopp, A. S. Ruetschi, D. Jaccard, M. Gabay, D. A. Muller, J. M. Triscone, and J. Mannhart, Science **317** (5842), 1196 (2007).
[5]   S. Thiel, G. Hammerl, A. Schmehl, C. W. Schneider, and J. Mannhart, Science **313** (5795), 1942 (2006).
[6]   C. Cen, S. Thiel, J. Mannhart, and J. Levy, Science **323** (5917), 1026 (2009).
[7]   D. F. Bogorin, C. W. Bark, H. W. Jang, C. Cen, C. M. Folkman, C. B. Eom, and J. Levy, Appl. Phys. Lett. **97** (1), 013102 (2010).
[8]   Patrick Irvin, Daniela F. Bogorin, Yanjun Ma, Cheng Cen, Jeremy Levy, and Chang-Beom Eom, arxiv:1009.2670v1 (2010).
[9]   Naoyuki Nakagawa, Harold Y. Hwang, and David A. Muller, Nat. Mater. **5** (3), 204 (2006).
[10]  Wolter Siemons, Gertjan Koster, Hideki Yamamoto, Walter A. Harrison, Gerald Lucovsky, Theodore H. Geballe, Dave H. A. Blank, and Malcolm R. Beasley, Phys. Rev. Lett. **98** (19), 196802 (2007).
[11]  G. Herranz, M. Basletic, M. Bibes, C. Carretero, E. Tafra, E. Jacquet, K. Bouzehouane, C. Deranlot, A. Hamzic, J. M. Broto, A. Barthelemy, and A. Fert, Phys. Rev. Lett. **98** (21), 216803 (2007).
[12]  Zoran S. Popovicacute, Sashi Satpathy, and Richard M. Martin, Phys. Rev. Lett. **101** (25), 256801 (2008).
[13]  K. Yoshimatsu, R. Yasuhara, H. Kumigashira, and M. Oshima, Phys. Rev. Lett. **101** (2) (2008).
[14]  Natalia Pavlenko and Thilo Kopp, arXiv:0901.4610v4 (2009).
[15]  U. Schwingenschlogl and C. Schuster, Euro. Phys. Lett. **81** (1), 17007 (2008).
[16]  C. Stephen Hellberg, unpublished.
[17]  Yanwu Xie, Christopher Bell, Takeaki Yajima, Yasuyuki Hikita, and Harold Y. Hwang, Nano Lett. **10** (7), (2010).
[18]  G. J. H. M. Rijnders, G. Koster, D. H. A. Blank, and H. Rogalla, Appl. Phys. Lett. **70** (1888 ) (1997).